\documentclass[preprint,showpacs,preprintnumbers,amsmath,amssymb]{revtex4}

\usepackage{graphicx}% Include figure files
\usepackage{dcolumn}% Align table columns on decimal point
\usepackage{bm}% bold math

\begin{document}
%\preprint{to be submitted to {\bf Phys. Rev. B}}
\bibliographystyle{aip}
%\bibliographystyle{apsrev} \draft

%%%%%%%%%%%%%%%%%%%%%%%%%%%%%%%%%%%%%%%%%%%%%%%%%%%%%%%%%%%%%%%%%%%%%

\title{Chemical doping-induced gap opening and spin polarization in graphene}

\author{I. Zanella$^1$, S. Guerini$^2$, S. B. Fagan$^3$,  J. Mendes Filho$^1$, A. G. Souza Filho$^1$\footnote{Corresponding author -  E-mail:
agsf@fisica.ufc.br}}
\affiliation{$^1$Departamento de F\'{\i}sica, Universidade Federal
do Cear\'a, CP 6030, CEP 60455-900, Fortaleza, CE, Brazil
\\$^2$Departamento de F\'{\i}sica, Universidade Federal do
Maranh\~ao,
65080-040, S\~ao Luis, MA, Brazil \\
$^3$\'Area de Ci\^encias Naturais e Tecnol\'ogicas, Centro
Universit\'ario Franciscano,  CEP 97010-032, Santa Maria, RS, Brazil
 }

\date{\today}
\vspace{5cm}

\begin{abstract}
By using first principles calculations we report a chemical doping induced gap in graphene.
The structural and electronic properties of CrO$_3$ interacting with graphene
layer are calculated using ab initio methods based on the
density functional theory. The CrO$_3$ acts as an electron acceptor modifying
the original electronic and magnetic properties of the graphene surface
through a chemical adsorption. The changes induced in the electronic
properties are strongly dependent of the CrO$_3$ adsorption site and for some sites it is possible to open a gap
in the electronic band structure. Spin polarization effects are also predicted for some adsorption configurations.

\end{abstract}


\maketitle

Nanostructured carbon-based materials exhibit remarkable electronic properties making them promising
materials to be used in a wide range of technological applications, including a possible
 carbon-based electronics in  a near future \cite{a1, a2}.
Graphene, a single atom thick layer, is considered the mother structure of all sp$^2$ nanostructured
carbon (such as fullerenes, carbon nanotubes and ribbons) and its electronic structure has been
theoretically investigated since 1947 \cite{a3}.
 A breakthrough in carbon science was the recent observation of graphene single layer that
 was obtained by micromechanical cleavage of graphite \cite{a4,a5}.
 The linear electronic band structure at the corner of the Brillouin zone where
 there is a band crossing between the conduction and valence bands cross is
responsible for striking physical properties not observed in any
other material such as ballistic transport \cite{a5}, and quantum
Hall effect at room temperature \cite{a6, a7} and typical
relativistic phenomena such as Berry\'s phase and Klein paradox
 as a consequence of massless Dirac Fermions.
The chemical stability, scalability and the complete compatibility of graphene with current
 semiconductor technology pave the way for the next generation of devices operating with a carbon based electronic.

The possibility of tailoring the electronic properties (type and number of carriers) of
 graphene is important not only for basic studies but also for further optimization of their
applications mainly as nanoscale based-sensors or spin filter
devices \cite{a8, a9, a10, a11, a12}. The electronic properties can
be tuning either by carrying out donor/acceptor doping experiments
or by applying a gate bias voltage \cite{a10, a13}.
 The electric field polarization and the electronegativity of the doping species modify
the carrier type (electron or hole) and carrier concentration at the
Dirac point affecting the electronic structure and consequently the
transport properties of graphene to a large extent.
 In this regard, a recent experimental work by Schedin et al. \cite{a10} fabricated and
operated a single molecule based sensor. The authors show, through
measurements of quantum Hall effect, that NO$_2$, H$_2$O and iodine
acts as electron acceptors whereas NH$_3$, CO and ethanol present
electron donor behavior on the graphene surface. This results were
 recently confirmed through theoretical
calculations \cite{a14}.
 Therefore, by studying the doping species in a controlled way, it is possible to access
 how specific chemical species perturb the graphene electronic band structure and how this
impact their physical properties such as magneto-optical and
transport behavior \cite{a10, a12}. In addition, the minimum
conductivity problem in graphene is a typical phenomenon of zero
band gap electronic  structure, and this has introduced drawbacks
for developing a graphene-based electronics. Carbon ribbons have
been proposed as one of the solution for the minimum conductivity
problem because in general its band structure exhibits a gap
\cite{a15}. The gap engineering can be also achieved through doping
(covalent functionalization) the graphene layer with chemical
species \cite{a16} since the doping can induce a gap opening because
of the breaking of translational symmetry. Therefore, the gap engineering in graphene is a hot
topic and this has been exploited either by changing the geometry of ribbons or by interaction with
substrate.\cite{a15,zhou07} The possibility of induce a gap in graphene through chemical doping
has not yet been exploited.

In this work, the electronic properties of a single graphene layer
interacting with CrO$_3$ molecules are analyzed through {\it ab
initio} based calculations and we predict that is possible to get both
gap opening and spin polarization by this functionalization process.
A single graphene layer is a unique
system regarding the surface to volume ratio because it is solely surface and this
opens up the possibility of using CrO$_3$ intercalated graphene systems for
the oxidation of primary alcohols. Our calculations shown that
CrO$_3$ molecule (a model for strongly oxidizing molecules)
chemically binds to graphene surface and the electronic properties
of the original graphene are modified by charge transfer from
graphene to this molecule leading the CrO$_3$ molecule act as an
electron acceptor with a similar behavior observed and predicted for
CrO$_3$ interacting with single-wall carbon nanotubes (SWNT)
\cite{a17}. The effect of curvature on the adsorption of this
molecule on carbon hexagonal lattice is also discussed.

First principles density functional theory has been employed to
investigate electronic and structural properties of graphene
interacting with CrO$_3$ molecule \cite{a18}. The SIESTA code is
used \cite{a19,a21},  which performs fully self-consistent
calculations solving the spin polarized Kohn-Sham equations
\cite{a20}. For the exchange and correlations terms,
 generalized gradient approximation with the parameterization of Perdew et al. is used \cite{a22}.
 The interaction between ionic cores and valence electrons is described by norm conserving
pseudopotentials \cite{a23} in the Kleinman-Bylander form \cite{a24}.
The Kohn-Sham orbitals are represented with a linear combination of pseudoatomic
orbitals with a double zeta basis set plus polarization function \cite{a21}.
A cutoff of 200 Ry for the grid integration is used to represent the charge density.
A 5x5x1 Monkhorst-Pack grid is employed for the Brillouin zone integration,
 which was shown to represent correctly their properties for the present system \cite{a25}.

    The periodic boundary conditions and a supercell approximation are used.
The supercell has 32 carbon atoms for a single graphene surface in a hexagonal lattice.
The structural optimizations were performed using a conjugated gradient procedure
and atomic positions of the structure are relaxed until all the force
components are smaller than 0.05 eV/$\AA$.

Different configurations of CrO$_3$ approaching to the graphene layer were
considered in our calculations and the most stable arrangements are shown
(top and side view) in Fig. 1(a)-(e). The configuration showed in Fig.1 (a)
 was predicted to be the most stable, with a Cr-C distance of about 2.32 $\AA$,
while in the configuration showed in Fig.1 (e) O-C bond length is
about 2.95 $\AA$. In Table 1 we list the closest distances Cr-C and
O-C for the different configurations showed in the Figure 1.

\begin{figure}
\label{fig1}
\includegraphics[angle=0,width=14.0 cm]{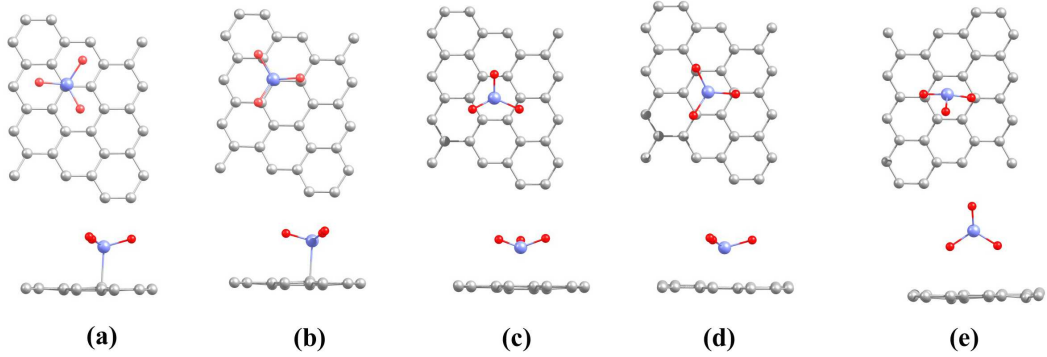}
\vspace{-14 cm} \caption{Top and side view of the most stable
configurations for the CrO$_3$ interacting with graphene layer.}
\end{figure}

The binding energies ($E_B$) for the studied systems are calculated
 using the basis set superposition error (BSSE) \cite{bsse}. This correction is
done through the counterpoise method using ''ghost" atoms, as the
following equation

\begin{equation}
\label{um}
 E_{B}= -[E_{T}(graphene+CrO_{3}) - E_{T}(graphene_{ghost} + CrO_{3}) - E_{T}( graphene + (CrO_{3})_{ghost})],
 \end{equation}
where $E_{T}(graphene+CrO_{3})$ is the total energy of the system.
The ``ghost" molecule/graphene corresponds to additional basis wave
functions centered at the position of the CrO$_{3}$ molecule or the
single layer graphene, but without any atomic potential.

\begin{table}[h]
 \caption{Binding energies ($E_B$), minimal distances and
 charge transfer calculated for different adsorption sites of CrO$_3$ on graphene layer
 as shown in Figure 1. The minus sign  in the charge transfer indicates that the  CrO$_{3}$
 molecule receive electronic charge.  }
\begin{center}
\begin{tabular}{ccccc}
\hline \hline Configuration $$& $E_B$  $$& d(Cr-C) $$& d(O-C)  $$ & Charge Transfer  \\
                            $$& (eV)   $$ &$\AA$ $$& $\AA$ $$ & to
                            CrO$_3$ ($e$) \\
 \hline
Fig. 1(a) & 1.01 & 2.32 & - & -0.17   \\
Fig. 1(b) & 0.91 & 2.33 & -  & -0.15 \\
Fig. 1(c) & 0.56 & 2.80 & - & -0.11  \\
Fig. 1(d) & 0.49 & 2.82 & - & -0.10  \\
Fig. 1(e) & 0.25 & - & 2.95 & -0.20  \\
\hline \hline
\end{tabular}
\end{center}
\end{table}

The binding energies obtained for the different adsorption sites of
CrO$_3$ on the graphene surface are listed in Table I. These results
pointed out to a  physical or chemical interaction between CrO$_3$
and graphene strongly  dependent  on the interaction site. The
transition metal atom (Cr) is preferentially bonded to the carbon
atom of the graphene layer promoting an $sp^3$-like hybridization of
the carbon atom from the graphene surface. The M\"ulliken population
was analyzed and used for all studied configurations to predict the
electronic charge transfer from the graphene surface to the CrO$_3$
molecule.
 The graphene carbon atoms transfer 0.17$e$ and 0.20$e$ for the CrO$_3$ molecule on the (Fig.1 (a)) and (Fig.1 (e)) configurations, respectively.
 The charge transfer on these cases occurs due to the chemical or physical
 adsorption
of the Cr or O atoms with the carbon atoms of the graphene layer as observed on the binding energies values.
 We should point out that the M\"ulliken population does not supply a trustworthy number
to charge transfer, but it indicates the trend and correct order of the charge transfer process.
 Then, the CrO$_3$ molecule behaves as an electron acceptor when interacting with
the graphene layer similar to what was observed for CrO$_3$
adsorption on SWNT surface \cite{a17}.

\begin{figure}
\label{fig2}
\includegraphics[angle=0,width=10.0 cm]{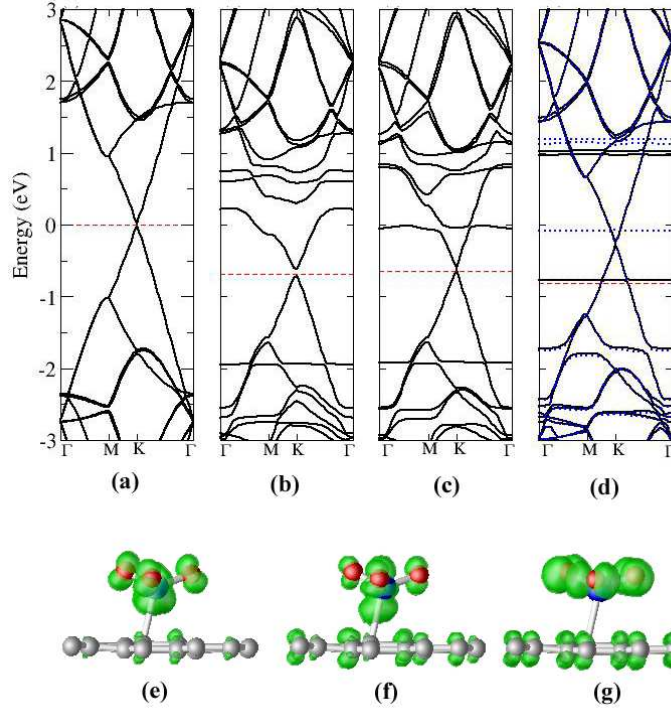}
\vspace{-4cm} \caption{Electronic band structures for (a) pristine
and for different configurations of CrO$_3$ adsorbed on graphene
single layer (b), (c) and (d) that correspond to the configurations
showed in Fig.1(a), (c), and (e), respectively. The (d) case has the
filled and dotted lines corresponding, respectively, to the majority
and minority spin levels. The horizontal dashed lines correspond to
the Fermi energy. The (e), (f) and (g) show the plots  for the
charge isosurface for CrO$_3$ interacting with graphene  in
electronic configuration presented in (b), for levels located around
0.8, -1.2 and -1.8 eV, respectively.}
\end{figure}

The calculated electronic band structures are presented for the pristine
single graphene layer (Fig. 2(a)) and for CrO$_3$ adsorbed on
graphene (Fig.2 (b), (c), and (d)) in the most stable configurations
showed in Figs.1 (a), (c) and (e)). It is clear that the CrO$_3$
 adsorption sites and orientation of the molecules affects the
electronic properties of graphene being more pronounced for the
 adsorption sites showed in (Fig.1 (a)). The binding of CrO$_3$ molecule lifts
the degeneracy of the band crossing close to the Fermi level, thus
inducing a gap opening of about 0.12 eV at the Dirac point, which
connects the valence and conduction bands at the zone edge. This gap
opening is due to the breaking of graphene mirror symmetry. For the
configuration shown in the Fig.1(b) the electronic behavior is
similar to the Fig.2(b). In contrast, for the other configurations
(Fig.1 (c)), Fig.1(d) and (Fig.1 (e)) the band crossing is preserved
leaving the system semimetal at 0 K. For the (Fig.1(c))
configuration the system remains metallic (see (Fig. 2c)), with
dispersionless levels of the CrO$_3$ molecule on the valence and
conduction bands. In the case (Fig.2 (d)) the majority (solid lines)
and minority (dotted lines)
 carriers coming from 3d-Cr levels are non-degenerated and located on the valence
and conduction band.  Comparing the adsorption we observe that the most
stable adsorption site changes the electronic properties due to the high
 hybridization between the molecule and graphene levels.
 It is also interesting to observe the downshift of the Fermi
 level resulting from the charge transfer from the graphene
to the CrO$_3$ molecule. This shift relative to the pristine
graphene is approximately 0.69, 0.64 and 0.8 eV showed in (Fig.
2(b)), (Fig. 2(c)), and (Fig. 2(d)), respectively. From all the
studied configurations only the site with the CrO$_3$ adsorbed on
the graphene surface, as showed on the (Fig. 1(e)), presents spin
polarization of 0.4 $\mu_{B}$. This spin polarization can be
understood in terms of redistribution of the electronic charge due
to the interaction of the O atoms of the CrO$_{3}$ molecule and the
graphene surface. In this case, a flat electronic level is located
near the Fermi level (Fig. 2(d)) that is characteristic of the
CrO$_3$ molecule.

In Fig. 2(e),(f) and (g) the plots for the localized density of
states (LDOS) of CrO$_3$ adsorbed on the graphene surface for the
most stable configuration (showed in Fig. 1(a)) are presented for
the electronic levels located about 0.8, -1.2 and -1.8 eV,
respectively. From these LDOS plots we observe the hybridization of
the molecule levels with the carbon levels from the graphene layer.
This hybridization state is reflected in both binding energies and
charge transfers values.

By Comparing the adsorption of the CrO$_3$ molecule on graphene
with adsorption on (8,0) SWNT surfaces \cite{a17}, it is observed that the curvature effect
is a crucial point to increase the adsorption of CrO$_3$ on the
carbon hexagonal lattice. The binding energy increase by 0.4 eV for
the molecule adsorbed on the inner or outer surface of the nanotube
(1.4 eV for both configurations) compared with the most stable
configuration on the graphene surface (1.01 eV). This effect is
relevant when is considered the real curved configuration (ripples)
 of the graphene layers \cite{a8}, which will increase the efficiency of the molecule
adsorption on the hexagonal carbon structures depending of the local
curvature parameters.

In summary, the electronic properties of the CrO$_3$ molecule
adsorbed on the graphene single layer were analyzed through first
principles calculations. It is observed that the electronic
properties and the charge transfer process are very sensitive to the
CrO$_3$ adsorption site and also to the orientation of the molecule.
This behavior is different from what was predicted Leenaerts et al.
\cite{a14} for other molecules, where the interaction is not very sensitive to the
adsorption site being only dependent on the molecule orientation
with respect
 to the graphene surface. The CrO$_3$ molecule interacts through a chemisorption or physisorption
regime on the graphene surface depending on the adsorption site,
indicating possible routes for the use doped graphene layer
 as platform for carbon-based electronics since it is possible to engineer the gap
and overcome the minimum conductivity problem.
Furthermore, the sensitivity of electronic structure towards doping makes
the graphene a potential material for chemical sensors owing to their
exceptionally high surface to volume ratio.

 {\vspace{0.2 cm}}
The authors thanks A. H. Castro Neto and B. Uchoa for the valuable
discussions. The authors also acknowledge CENAPAD-SP for computer
time and financial support from Brazilian agencies CNPq, FAPERGS
(Grants 0511570/05 and 05/2096.2) and FUNCAP (Grant 350220/2006-9).
S.B.Fagan acknowledges the Brazilian Woman in Science/2006 grant
from L´oreal/Paris.

\end{document}